\documentclass{aa}
\usepackage{graphicx}
\usepackage{txfonts}
\usepackage{natbib}

\begin{document}
  \headnote{Research Note} 
  
  \title{Celestial position of the companion of PSR~J1740$-$5340}

  \author{C.\ G.\ Bassa\inst{1}
    \and B.\ W.\ Stappers\inst{2,3}}

  \institute{
    Astronomical Institute, Utrecht University, PO Box 80\,000, 3508
    TA Utrecht, The Netherlands 
    \and Stichting ASTRON, Postbus 2, 7990 AA Dwingeloo, The Netherlands
    \and Astronomical Institute ``Anton Pannekoek'', University of
    Amsterdam, Kruislaan 403, 1098 SJ Amsterdam, The Netherlands
  }

  \offprints{C.\ G.\ Bassa,\\ \email{c.g.bassa@astro.uu.nl}}

  \date{Received  11 May 2004 / Accepted 20 July 2004}

  \abstract{ 
    We present optical astrometry of archival ground and space based
    imaging of the companion to PSR~J1740$-$5340. The optical position
    of the companion is significantly offset from the timing position
    of the pulsar. We briefly investigate the effects of this
    inconsistency on other timing parameters and compare our position
    with an improved position of PSR~J1740$-$5340 from recent,
    preliminary, timing results.
    \keywords{Pulsars: individual (\object{PSR~J1740$-$5340}) 
      -- astrometry
      -- globular clusters: individual (\object{NGC\,6397})}}

  \maketitle

  \section{Introduction}
  In many ways, the binary millisecond pulsar PSR~J1740$-$5340
  \citep{dpm+01}, located in the nearby (2.3~kpc) galactic globular
  cluster NGC\,6397, is an exceptional system; it shows irregular
  eclipses over a wide range of orbital phases and it has the largest
  orbital period and the heaviest binary companion among all currently
  known eclipsing pulsars, including those in globular
  clusters\footnote{For an up-to-date list, see
  http://www.naic.edu/\~{}pfreire/GCpsr.html}. An optical variable,
  discovered by \citet{tgec01}, was identified as the pulsar companion
  by \citet{fpds01} on the basis of positional coincidence with the
  timing position of the pulsar. These and later photometric and
  spectroscopic observations \citep{ok03,krt03,fsg+03} have revealed
  that both the brightness and the radial velocity of the companion
  vary with the orbital period and phase of the pulsar, unambigously
  linking this star as the companion to PSR~J1740$-$5340.

  The observed pulsar period $P$ and period derivative $\dot{P}$ give
  rise to a large spin-down luminosity, $L_\mathrm{SD}\propto
  \dot{P}/P^3\approx1.4\times10^{35}$~erg~s$^{-1}$ \citep{dpm+01},
  which is among the highest found for millisecond pulsars. Though
  part of the period derivative may not be intrinsic, due to
  accelerations in the potential of the globular cluster, the effects
  of irradiation of the companion by the pulsar would give rise to
  heating. No evidence of heating of the pulsar companion, at the
  level expected from the spin-down luminosity, are found
  \citep{ok03}.

  From accurate absolute optical astrometry of \emph{HST}/WFPC2
  observations of NGC\,6397 we have found that the optical position of
  the companion of PSR~J1740$-$5340 is inconsistent with the position
  of the pulsar \citet{dpm+01}. A preliminary timing solution
  (A.~Possenti, priv.~comm.) provides a celestial position that is in
  much better agreement with our optical position, but moreover,
  predicts a significantly different spin-down luminosity, now in
  agreement with the observed lack of heating of the pulsar
  companion. 

  \section{Observations and data reduction}
  The method that was employed to determine the optical position of
  the companion of PSR~J1740$-$5340 is similar to that described in
  \citet{bvkh03}; we will focus on the differences. We used an
  archival 4~minute $V$-band exposure, obtained with the Wide Field
  Imager (WFI) at the ESO 2.2~m telescope at La Silla{\bf ,} during
  the night of May 14, 1999. A total of 248 stars from the UCAC2, the
  2nd version of the USNO CCD Astrograph Catalog\footnote{UCAC,
  http://ad.usno.navy.mil/ucac/} \citep{zuz+04}, coincident with an
  $8\arcmin\times8\arcmin$ subsection of the chip containing the
  cluster center, were selected for the astrometry and their positions
  were measured. Of these, 88 stars were not saturated and appeared
  stellar and unblended. An astrometric solution, fitting for
  zero-point position, scale and position angle, was computed and 6
  outliers, having residuals in excess of $0\farcs25$, were
  iteratively removed. The final solution of 82 stars had rms
  residuals of $0\farcs068$ in both coordinates.
  \begin{figure}
    \resizebox{\hsize}{!}{\includegraphics[angle=270]{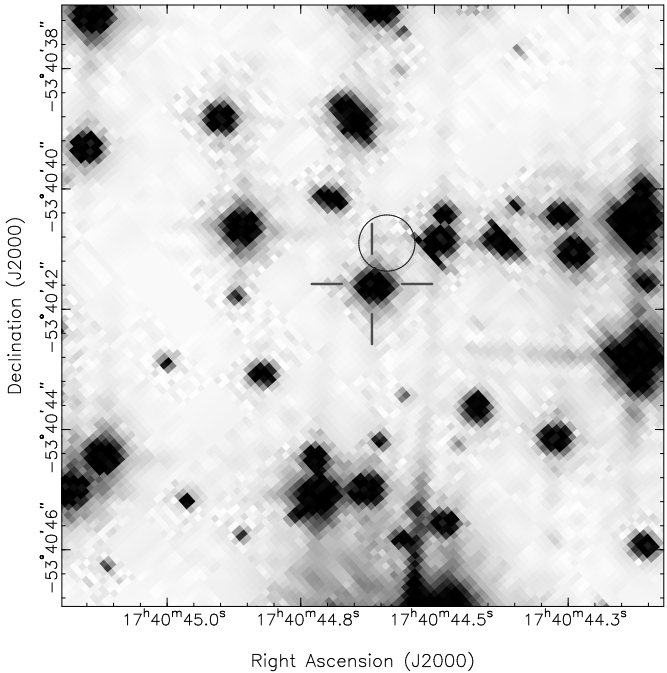}}
    \caption{A $10\arcsec \times 10\arcsec$ subsection of a 249~s
    $V_{555}$ image from the \emph{HST}/WFPC2 GO5929 dataset. The 99\%
    confidence level on the position of PSR~J1740$-$5340 by
    \citep{dpm+01} is indicated with a circle. The radius of the
    circle is $0\farcs47$ and incorporates the uncertainties in the
    pulsar position, in the astrometric tie between the WFI and the
    UCAC2 and in the tie between the WFI and the \emph{HST}/WFPC2
    image. The position of the pulsar companion, as identified by
    \citet{fpds01} is indicated with tickmarks ($0\farcs5$ in
    length).}
    \label{fig:hst}
  \end{figure}

  This solution was transferred to F555W (hereafter $V_{555}$) images
  of \emph{HST}/WFPC2 datasets GO5929 and GO7335. As the pulsar
  companion is bright ($V_{555}=16.9$), we only used 4 $V_{555}$
  images, with a total exposure time of 89~s, from both
  datasets. Positions of stars on these images were obtained with the
  HSTphot 1.1.5b package \citep{dol00a}. We matched stars that were
  common to the \emph{HST}/WFPC2 datasets and the WFI image and fitted
  for zero-point position, scale and position angle between the WFI
  pixel positions and the distortion corrected \emph{HST}/WFPC2 master
  frame positions \citep{ak03}. Outliers were iteratively removed
  until the astrometric solution converged. The final astrometric
  solution contained 154 (190) stars, resulting in rms residuals of
  $0\farcs045$ ($0\farcs056$) in right ascension and $0\farcs050$
  ($0\farcs061$) in declination for the GO5929 (GO7335) dataset.

  The position of the optical companion to PSR~J1740$-$5340 in the WFI
  image and the two \emph{HST}/WFPC2 datasets is given in
  Table~\ref{tab:pos} while Fig.~\ref{fig:hst} shows a finding
  chart. The uncertainty in the position of the companion in the WFI
  image is the quadratic sum of the uncertainty in the tie between the
  UCAC2 and the WFI and the positional uncertainty of the companion,
  about $0\farcs06$. In the \emph{HST}/WFPC2 datasets the positional
  uncertainty of the companion is much smaller, and hence the
  uncertainty is dominated by that of the tie between the UCAC2 and
  the WFI and that of the tie between the WFI and the \emph{HST}/WFPC2
  images.

  The astrometry provided in the UCAC2 is on the Hipparcos system,
  i.e. the International Celestial Reference System (ICRS), on which
  the pulsar timing observations are based. Random positional
  uncertainties of the UCAC2 stars range from $0\farcs020$ for 10 to
  14th magnitude stars upto $0\farcs070$ for 16th magnitude stars
  \citep{zuz+04}. These uncertainties are incorporated in the
  uncertainty in the tie between the UCAC2 and the WFI. A comparison
  of UCAC positions with the International Celestial Reference Frame
  (ICRF), the primary representation of the ICRS \citep{fm98}, is
  presented in \citet{azrz+03}.

  \section{Discussion and conclusions}
  The timing position of PSR~J1740$-$5340 as published by
  \citet{dpm+01} is inconsistent with the optical position of the
  pulsar companion. The timing position,
  $\alpha_\mathrm{J2000}=17^\mathrm{h}40^\mathrm{m}44\fs589(4)$,
  $\delta_\mathrm{J2000}=-53\degr40\arcmin40\farcs9(1)$ \citep{dpm+01}
  is offset from the optical position by $-0\farcs23(8)$ and
  $0\farcs69(12)$ in right ascension and in declination, respectively.
  We attempted to use this position to fit simulated TOAs generated
  using the ephemeris of D'Amico~et~al. but found that even with quite
  large changes in the period derivative it was not possible to get a
  good fit of the TOAs. This indicated that there was perhaps
  something wrong with the exisiting timing solution.

  \begin{table}
    \centering
    \caption[]{Celestial position of the companion to
    PSR~J1740$-$5340.}\label{tab:pos}
    \begin{tabular}{
	l@{\hspace{0.2cm}}
	l@{\hspace{0.2cm}}
	l@{\hspace{0.2cm}}
	l@{\hspace{0.2cm}}
      }
      \hline \hline Dataset & Date (UT) &
      \multicolumn{1}{c}{$\alpha_\mathrm{J2000}$} &
      \multicolumn{1}{c}{$\delta_\mathrm{J2000}$} \\ 
      \hline 
      WFI & May 14, 1999 & $17^\mathrm{h}40^\mathrm{m}44\fs611(10)$ &
      $-53\degr40\arcmin41\farcs57(9)$ \\ 
      GO5929 & March 6--7, 1996 &  $17^\mathrm{h}40^\mathrm{m}44\fs617(9)$ &
      $-53\degr40\arcmin41\farcs58(8)$ \\ 
      GO7335 & April 3, 1999 &  $17^\mathrm{h}40^\mathrm{m}44\fs617(10)$ &
      $-53\degr40\arcmin41\farcs62(9)$ \\[0.2ex]
      \multicolumn{2}{l}{Mean:} & $17^\mathrm{h}40^\mathrm{m}44\fs615(8)$ &
      $-53\degr40\arcmin41\farcs59(7)$ \\
      \hline
    \end{tabular}
  \end{table}
  
  Recent timing results, using a longer data span, also revealed the
  inconsistent position (A.~Possenti, private communication). It was
  found that, as the radio signal of PSR~J1740$-$5340 is eclipsed at
  phases far from inferior conjunction, the reliability of TOAs is
  hard to assess, and, as a result, the errors appeared more uncertain
  than those quoted in the \citet{dpm+01} ephemeris. Improved
  astrometry (Possenti et al., in preparation) is in much better
  agreement with our optical position; a preliminary position by
  \citet{pdc+04} is offset from the position of the companion by
  $0\farcs13(7)$ in right ascension and $-0\farcs16(9)$ in
  declination. Though still outside the $1\sigma$ errors the accurate
  optical position of the companion might aid future timing of
  PSR~J1740$-$5340.

  As a result of the change in position, the preliminary ephemeris
  \citep{pdc+04} has an updated pulse period derivative that is
  significantly different from that of \citet{dpm+01}. Hence, the
  spin-down luminosity of PSR~J1740$-$5340 is decreased by over a
  factor 5; $L_\mathrm{SD}\approx3.3\times10^{34}$~erg~s$^{-1}$. This
  value is in much better agreement with the lack of heating of the
  companion by the pulsar as expected by \citet{ok03}.

  \begin{acknowledgements}
    We thank Andrea Possenti and Nichi D'Amico for sharing unpublished
    results. CGB acknowledges support by the Netherlands Organization
    for Scientific Research.
  \end{acknowledgements}

  \bibliographystyle{aa}
  
\end{document}